\documentclass[prl,twocolumn,aps,showpacs,floatfix,superscriptaddress]{revtex4}

\usepackage{graphicx}
\usepackage{rotating}
\usepackage{amsmath}
\usepackage{bbm}
\usepackage{subfigure}
\usepackage{color}
\usepackage{natbib}

\newcommand{\be}{\begin{equation}}
\newcommand{\ee}{\end{equation}}
\newcommand{\bea}{\begin{eqnarray}}
\newcommand{\eea}{\end{eqnarray}}

\begin{document}

\title{Density functional theory beyond the linear regime: Validating adiabatic LDA}
\author{N.~Helbig}
\author{J.~I.~Fuks}
\affiliation{Nano-Bio Spectroscopy group and ETSF Scientific Development Centre, Dpto.\ F\'isica de Materiales, Universidad del Pa\'is Vasco, CFM CSIC-UPV/EHU-MPC and DIPC, Av.\ Tolosa 72, E-20018 San Sebasti\'an, Spain}

\author{M.~Casula}
\affiliation{CNRS and Institut de Min\'eralogie et de Physique des Milieux condens\'es,
case 115, 4 place Jussieu, 75252, Paris cedex 05, France}

\author{M.~J.~Verstraete}
\affiliation{Institut de Physique, Universit\'e de Li\`ege, B-4000 Sart Tilman, Belgium}
\affiliation{European Theoretical Spectroscopy Facility}

\author{M.~A.~L.~Marques}
\affiliation{Laboratoire de Physique de la Mati\`ere Condens\'ee et Nanostructures,
Universit\'e Lyon 1 et CNRS, F-69622 Villeurbanne Cedex, France}
\affiliation{European Theoretical Spectroscopy Facility}

\author{I.~V.~Tokatly}
\affiliation{Nano-Bio Spectroscopy group and ETSF Scientific Development Centre, Dpto.\ F\'isica de Materiales, Universidad del Pa\'is Vasco, CFM CSIC-UPV/EHU-MPC and DIPC, Av.\ Tolosa 72, E-20018 San Sebasti\'an, Spain}
\affiliation{IKERBASQUE, Basque Foundation for Science, E-48011 Bilbao, Spain}

\author{A.~Rubio}
\affiliation{Nano-Bio Spectroscopy group and ETSF Scientific Development Centre, Dpto.\ F\'isica de Materiales, Universidad del Pa\'is Vasco, CFM CSIC-UPV/EHU-MPC and DIPC, Av.\ Tolosa 72, E-20018 San Sebasti\'an, Spain}
\affiliation{Fritz-Haber-Institut der Max-Planck-Gesellschaft, Faradayweg 4-6,
D-14195 Berlin, Germany}

\begin{abstract} 
We present a local density approximation (LDA) for one-dimensional (1D) systems
interacting via  the soft-Coulomb interaction based on
quantum Monte-Carlo calculations.  Results for the ground-state energies and
ionization potentials of finite 1D systems show excellent agreement with
exact calculations, obtained by exploiting the mapping  of an $N$-electron
system in $d$ dimensions, onto a single electron in $N\times d$ dimensions
properly symmetrized by the Young diagrams. We conclude that 1D LDA is of the
same quality as its three-dimensional (3D) counterpart, and we infer
conclusions about 3D LDA. The linear and non-linear time-dependent
responses of 1D model systems using LDA, exact exchange, and the exact solution
are investigated and show very good agreement in both cases, except for the well
known problem of missing double excitations. Consequently, the 3D LDA is
expected to be of good quality beyond linear response. In addition, the 1D LDA
should prove useful in modeling the interaction of atoms with strong laser
fields, where this specific 1D model is often used.
\end{abstract}

\date{\today}

\maketitle

Over the last years the theoretical description of optical properties of
complex many-electron systems, from molecules, to nanostructures and extended
systems, has re-flourished due to the efficient implementation of time-dependent
density-functional theory (TDDFT) \cite{RG1984,TDDFT2006}. The good performance
shown by the adiabatic local density approximation (ALDA) for many finite
systems has limited the development of exchange-correlation (xc) functionals
with a more elaborate time-dependence, which is clearly in its infancy compared
to static DFT. However recently many important deficiencies, especially of the
adiabatic approximation, have been identified
\cite{ORR2002,MBW2002,MZCB2004,DH2004,M2005,BWG2005,VMR2008,CCJ2009,GMG2009,GB2009}.

Ultrafast time-resolved optical spectroscopy has revealed new classes of
physical, chemical, and biological reactions, in which directed, deterministic
motions of atoms have a key role. The advent of free electron lasers with
attosecond resolution increases the capabilities of present femtosecond
pump-probe experiments, allowing for a study of the dynamics of non-equilibrium
electronic  systems in real time.  In addition, systems of all sizes can be investigated, from  the
atomic scale to the most extended molecules (e.g. DNA, proteins and their
complexes) and solids. Despite those tremendous
experimental advances, the theoretical description of a real molecular system
subject to ultrashort, intense, and/or high-frequency lasers is still in  a
fledgeling state. Several problems need to be addressed, ranging from the
non-perturbative nature of the physical processes involved to the simultaneous
description of the (interacting) electronic and nuclear degrees of freedom.
Therefore, it is of paramount relevance to have a theoretical framework which
allows for a non-perturbative description of electrons, and at the same time
is able to tackle electron-ion dynamics in the excited-state. TDDFT seems to be
the suitable framework to move the realm of density-functional methods  beyond
the linear regime to describe the aformentioned processes. One important
advantage is  the combined electron and ion dynamics provided by TDDFT
\cite{TDDFT2006}. 

Many physical processes rely on the knowledge of  non-linear response
functions. Therefore, it is very timely to provide a systematic study
addressing the performance of present functionals in the non-linear regime.
To assess the quality of a functional we need to have appropriate data
for comparison. Obtaining accurate experimental data in the non-linear regime
can be very difficult for real systems, due to various limitations, e.g. solvent
effects or additional approximations going into the interpretation of the
collected data \cite{SMBWRSD1993,KDS1998}. These problems can be avoided by
using exactly solvable  models which then  allow for a direct comparison
between the exact spetrum and an approximate one. Unfortunately, an exact
propagation of even small three-dimensional systems is computationally very
demanding,  and needs further simplification. One possibility is the reduction
of dimensionality, i.e. the use of one-dimensional (1D)  models,  where the exact
diagonalization is feasible as long as the number of electrons is sufficiently small.
In the present paper we
work with systems of interacting electrons in 1D. Having the exact
solution allows us to test orbital dependent functionals such as exact exchange
(EXX) which can be easily transferred to different dimensions. A local density
approximation (LDA) is achieved, as in the 3D case, by quantum Monte-Carlo
(QMC) studies  of the homogeneous reference, and parametrizing the
corresponding correlation energy. 

The present work, besides adding fundamental information concerning the
relevance of spatial and time non-locality in the xc functional, also provides 
a proper LDA parametrization for electrons interacting via the soft-Coulomb
interaction in 1D systems. This model description is widely used in the context
of high-intensity lasers, where above-threshold ionization and high-harmonic
generation play an important role \cite{JES1988,E1990,LGE2000,L2005}. Also, 1D
two-electron systems are employed to gain insight into exact properties of the
xc potential and kernel in static and time-dependent density functional theory,
since these systems can easily be solved exactly \cite{TGK2008,TMM2009,HTR2009}.

The 1D Hamiltonian for $N$ particles moving in a general external potential $v_{ext}$ reads
\be\label{eq:1dham}
H=\sum_{j=1}^N \left[ -\frac{1}{2} \frac{d^2}{dx_j^2}+v_{ext}(x_j)\right]+\frac{1}{2}
\sum_{\stackrel{\scriptstyle j,k=1}{j\neq k}}^N v_{int}(x_j, x_k),
\ee
where $v_{int}$ describes the electron-electron interaction. In order to avoid
the singularity of the Coulomb interaction  we employ the soft-Coulomb
potential
\be\label{eq:softC}
v_{soft-C}(x_1, x_2) = \frac{q_1q_2}{\sqrt{a^2+(x_1-x_2)^2}}
\ee
instead. Here, $q_1$ and $q_2$ describe the charges of the particles while $a$
is the usual softening parameter (atomic units $e=m=\hbar=1$ are used
throughout this paper). Unless stated explicitly, we use $a=1$ for all our
calculations. Mathematically, it is straightforward to show that the
Hamiltonian (\ref{eq:1dham}) is equivalent to a single particle in $N$
dimensions, moving in an external potential consisting of all the contributions
from $v_{ext}$ and $v_{int}$. The corresponding Schr\"odinger equation can,
hence, be solved by any code which is able to treat non-interacting particles
in the correct number of dimensions in an arbitrary external potential. Due to
the Hamiltonian being symmetric under particle interchange, $x_j\leftrightarrow
x_k$, the solutions of the Schr\"odinger equation can be chosen as symmetric or
antisymmetric under such an exchange. For the simplest case of two interacting
electrons both the symmetric and antisymmetric solutions are valid,
corresponding to the singlet and triplet spin configurations, respectively.
However, for more than two electrons one needs to separately ensure that the
spatial wave function is a solution to the $N$-electron problem. For example, a
totally symmetric spatial wave function is a correct solution for a single
particle in $N$ dimensions, however, for $N>2$ there is no corresponding spin
function such that the total wave function has the required antisymmetry to be
a solution of the $N$ fermion problem in 1D. We solve this problem by
symmetrizing the solutions according to all possible fermionic Young diagrams
for the given particle number $N$ \cite{LL1977}. The solution of higher
dimensional problems within these symmetry restrictions has been implemented
into the {\tt OCTOPUS} computer program \cite{MCBR2003,CAORALMGR2006}. Usually,
the lowest energy solution is found to be purely symmetric and is discarded
for $N>2$. With increasing number of electrons we also observe an
increasing number of states which do not satisfy the fermionic symmetry
requirements.

As a result of reducing the number of dimensions, we need to use an appropriate
functional for performing the DFT calculations. While any orbital functional
can easily be transferred between dimensions, those functionals based on
specific systems need to be recalculated. This affects the most common
functional,  i. e. the local density approximation,  available  only  for the
normal Coulomb interaction in two and three dimensions \cite{AMGB2002,KS1965},
an effective Coulomb interaction of a  harmonically confined wire
\cite{CSS2006,SCSM2009}, and some other \emph{ad-hoc} 1D models
\cite{XPTCCR2006,XPAT2006,XA2008}. In this work, we present and use a
parametrization of the 1D LDA obtained from quantum Monte-Carlo simulations,
which are exact in 1D,  using the soft-Coulomb interaction  in
Eq.~\ref{eq:softC}. We assess the quality of the approximation in calculating
ground-state properties as well as the linear response for various 1D model
systems. We then proceed to calculate the nonlinear response and compare the
exact one with the ALDA and  adiabatic exact-exchange (AEXX)  spectra. 

The correlation energy of the LDA is parametrized in terms of  $r_s$ and the spin polarization $\zeta=(N_\uparrow-N_\downarrow)/N$ in the form
\be
\label{eq:ectotal} 
\epsilon_c(r_s,\zeta)=\epsilon_c(r_s,\zeta=0)+\zeta^2
\left[\epsilon_c(r_s,\zeta=1)-\epsilon_c(r_s,\zeta=0)\right]
\ee
with
\bea
\nonumber
\epsilon_c(r_s,\zeta=0,1)&=&-\frac{1}{2}\frac{r_s + E r_s^2}{A+ B r_s
+C r_s^2 + D r_s^3 } \\[2ex]
\label{eq:ec} 
&&\times\ln ( 1 + \alpha r_s + \beta r_s^m ), 
\eea 

which proved to be very accurate in the parametrization for other 1D systems
with a different long-range interaction \cite{CSS2006}. Note that the additional
factor of $1/2$ is due the use of Hartree atomic units,
as everywhere else in the paper. To obtain 
the exact high-density result, known from  the random phase approximation
\cite{SCSM2009}, i.e. 
\bea
\epsilon_c(r_s \rightarrow 0,\zeta=0) &=& -4/(\pi^4 a^2)~r_s^2,\\
\epsilon_c(r_s \rightarrow 0,\zeta=1) &=& -1/(2 \pi^4 a^2)~r_s^2
\eea
to leading order in  $r_s$, we fix the ratio  $\alpha/A$ to be equal to
$8/(\pi^4 a^2)$ and $1/(\pi^4 a^2)$ for $\zeta=0$ and $\zeta=1$, respectively. 
In both cases the exponent $m$ is limited to values larger than 1. As a result, the number
of independent parameters in  the function (\ref{eq:ec}) is reduced to 7. In
addition, for $a=1$ the denominator can be simplified by setting $B=0.0$.
However, for smaller values of the softening parameter $a$ the linear term in
the denominator is important for achieving better agreement with the quantum
Monte Carlo results. 
\begin{table}
\begin{tabular}{|l||c|c||c|}\hline
    & \multicolumn{2}{|c||}{$a=1.0$} & \multicolumn{1}{|c|}{$a=0.5$} \\
    & $\zeta=0$ & $\zeta=1$ & $\zeta=0$ \\ \hline \hline
$A$      & 18.40(29)    & 5.24(79)                   & 7.40(18)  \\
$B$      & 0.0          & 0.0                        & 1.120(119)  \\
$C$      & 7.501(39)    & 1.568(230)                 & 1.890(63) \\
$D$      & 0.10185(5)   & 0.1286(150)                & 0.0964(108) \\
$E$      & 0.012827(10) & 0.00320(74)                & 0.0250(23) \\
$\alpha$ & 1.511(24)    & 0.0538(82)                 & 2.431(62) \\
$\beta$  & 0.258(6)     & $1.56(1.31)\cdot 10^{-5}$  & 0.0142(25) \\
$m$      & 4.424(25)    & 2.958(99) &  2.922(83)  \\ \hline 
av. error &  $6.7\cdot 10^{-5}$ & $3.3\cdot 10^{-5}$ & $7.7\cdot 10^{-4}$ \\\hline
\end{tabular}
\caption{\label{tab:encorr_parameters} Values of the LDA correlation energy parametrization in Eq.~\ref{eq:ec}. For the most widely used case, i.e. $a=1$, the parametrization is reported for both unpolarized ($\zeta=0$) and fully polarized ($\zeta=1$) systems. The error on the last digits is given in parenthesis, while the average error (in Hartree) in the full density range is given in the last row.}
\end{table}  
The optimal values of the parameters for $a=1$ and $a=0.5$ are reported in
Tab.~\ref{tab:encorr_parameters}, and implemented in the {\tt
OCTOPUS} program \cite{MCBR2003,CAORALMGR2006}. For more details on the 1D QMC methodology and the parametrization procedure we refer to Refs.~\onlinecite{CSS2006,SCSM2009}. 

\begin{table}
\begin{tabular}{|l||c|c|c||c|c|c|}\hline
& \multicolumn{3}{|c||}{$E_{\mathrm{total}}$} & \multicolumn{3}{|c|}{IP} \\
& Exact &  LDA  &  SLDA & Exact&(S)LDA&  $\epsilon_{\mathrm{HOMO}}^{\mathrm{(S)LDA}}$ \\ \hline \hline
H         & -0.67 & -0.60 & -0.65 & 0.67 & 0.65 & -0.41 \\
He        & -2.24 & -2.20 &       & 0.75 & 0.75 & -0.48 \\
Li        & -4.21 & -4.16 & -4.18 & 0.31 & 0.33 & -0.18 \\
Be        & -6.78 & -6.76 &       & 0.33 & 0.35 & -0.16 \\
He$^+$    & -1.48 & -1.41 & -1.45 & 1.48 & 1.45 & -1.18 \\
Li$^+$    & -3.90 & -3.85 &       & 1.56 & 1.55 & -1.24 \\
Be$^+$    & -6.45 & -6.39 & -6.41 & 0.83 & 0.85 & -0.63 \\
Li$^{2+}$ & -2.34 & -2.25 & -2.30 & 2.34 & 2.30 & -2.00 \\
Be$^{2+}$ & -5.62 & -5.56 &       & 2.41 & 2.38 & -2.06 \\ 
Be$^{3+}$ & -3.21 & -3.13 & -3.18 & 3.21 & 3.18 & -2.86 \\\hline
\end{tabular}
\caption{\label{tab:totenergies}Total energies and ionization potentials for one-dimensional atoms and ions from exact and (spin-)LDA calculations as well 
as the eigenvalues of the highest occupied Kohn-Sham orbital. All numbers are given in Hartree.} 
\end{table}

As a first test, we calculated the ground-state energies of small atomic
systems, for example, a 1D helium atom with $q=2$ in Eq.~(\ref{eq:softC}) and
two electrons which interact via the soft Coulomb interaction. The ground state
energies and ionization potentials from the exact and unpolarized LDA
calculations are given in Tab.~\ref{tab:totenergies}. We include all possible
systems with one, two, three and four electrons in our test. For open-shell
systems, we additionally performed a spin-DFT (SLDA) calculation, where the xc
energy was spin dependent according to Eq.~(\ref{eq:ectotal}). All atomic
calculations were performed in a box ranging from -8 to 8~bohr with a spacing
of 0.2~bohr, which ensures the total energy to be converged to the accuracy
stated in the table. 

As we can see, the LDA total energies for the neutral and positively charged
systems agree very well with the exact results. As expected, the spin-resolved
calculations further improve the agreement for the open-shell systems. As a
result, the ionization potentials, calculated as the difference of the total
energies of the $N$ and the $N-1$ electron systems, from the (S)LDA and the
exact calculations agree almost perfectly. The largest Kohn-Sham eigenvalue
$\epsilon_{HOMO}^{(S)LDA}$ only partially accounts for the total ionization
potential, i.e. the 1D LDA violates this known property of the exact functional
\cite{AB1985b}. The good agreement for the positively charged systems is not
reproduced for negatively charged ones. For the small systems investigated
here, LDA does not bind an extra electron while the exact calculation shows
that the negatively charged systems are indeed stable giving total energies of
$-0.73$~Ha, $-2.35$~Ha and $-4.17$~Ha for H$^-$, He$^-$, and Li$^-$,
respectively. A comparison with the total energies of the neutral systems shows
that in the exact calculation the additional electron is only very lightly
bound in 1D. It is no surprise that the LDA, with its usual wrong
asymptotic behavior of the exchange-correlation potential, does not yield stable
negatively charged ions.

\begin{figure}
\includegraphics[width=0.47\textwidth,clip]{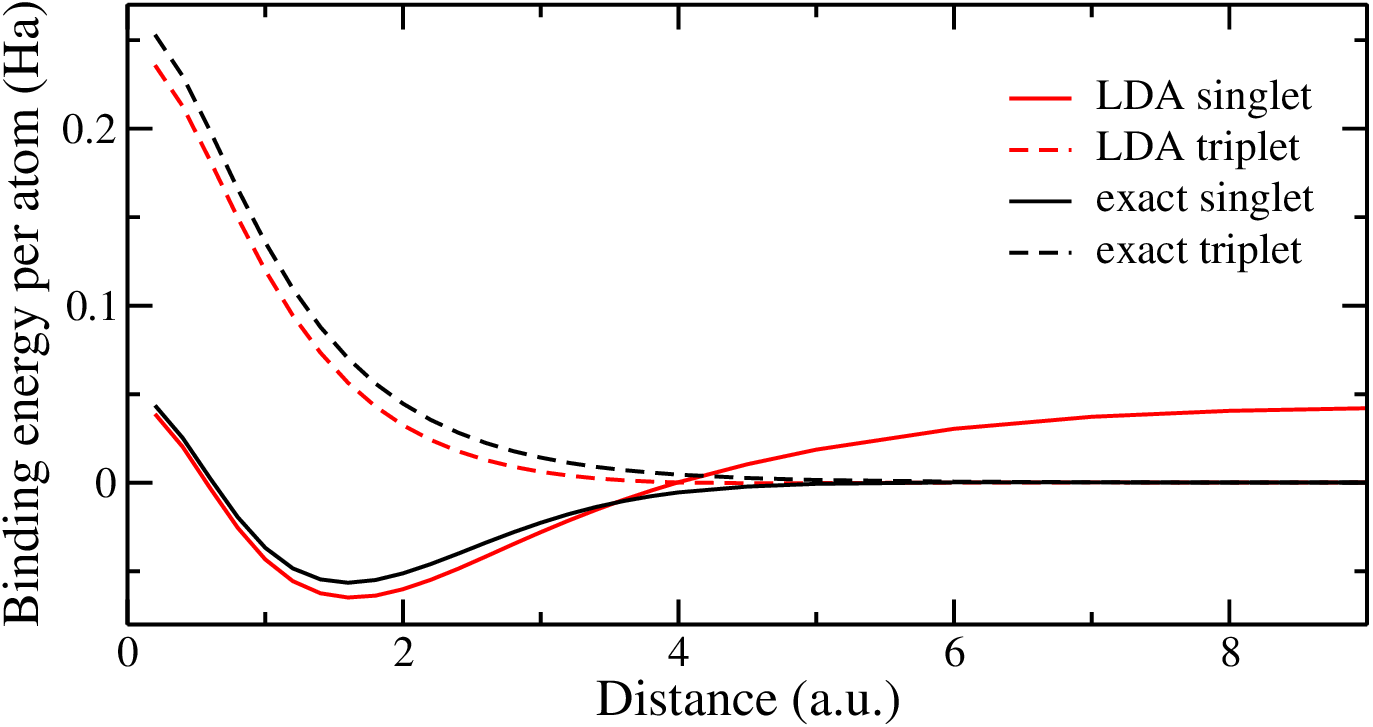}
\caption{\label{fig:h2diss} Binding energy per atom of the one-dimenstional hydrogen molecule as a function of the distance between the two ions, exact and LDA calculations for the singlet ground state and the first triplet state.}
\end{figure}

As a second test of the new functional we calculate the dissociation curve of
the 1D hydrogen molecule. For these calculations we increased the size of the
simulation box to range from -20 to 20~bohr in order to achieve convergence
also for the stretched molecule. Fig.~\ref{fig:h2diss} shows the binding energy
per atom as a function of the distance between the two ions. As one can see,
the known pathology of 3D LDA is reproduced also in 1D. The singlet state yields
a good description around the equilibrium distance of $1.6$~bohr with the
binding energy being slightly overestimated by LDA. However,  the bond
breaking  is not described correctly due to the strong static correlation at
large distances. The LDA calculation for the triplet state yields very good
agreement over the whole range of distances corroborating the general
experience of LDA performing better for more polarized systems.

\begin{figure}
\includegraphics[width=0.47\textwidth,clip]{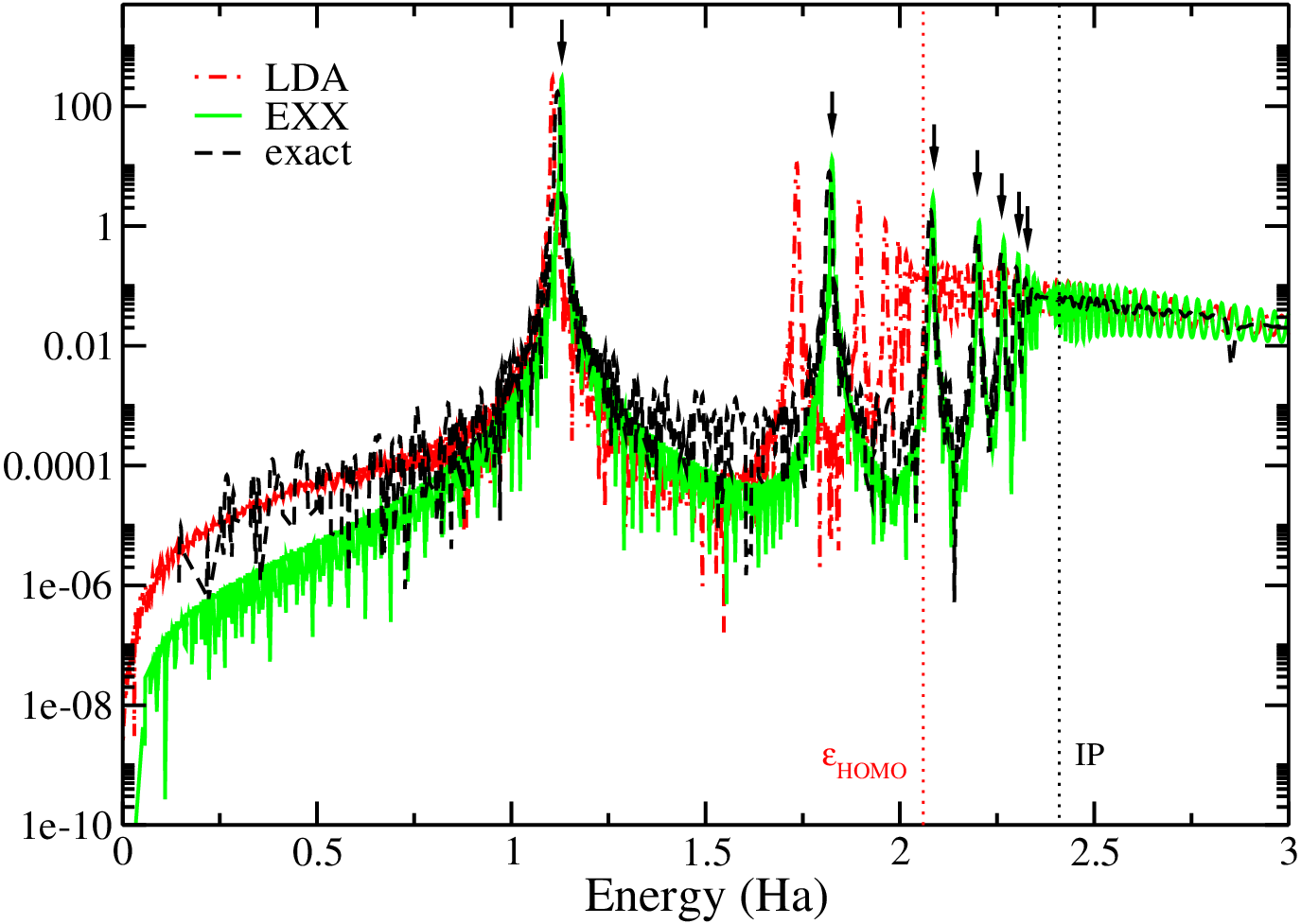}
\includegraphics[width=0.47\textwidth,clip]{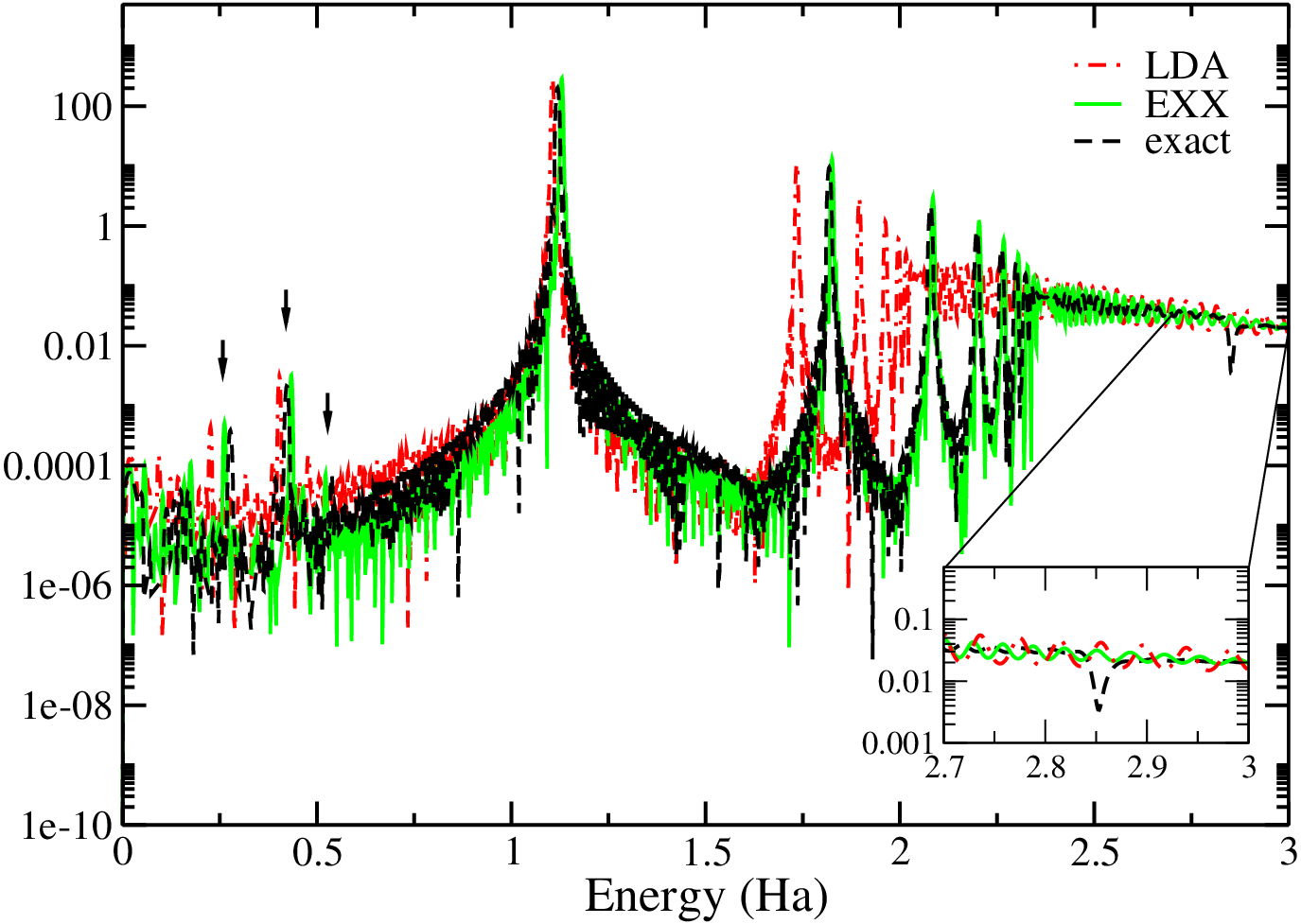}
\caption{\label{fig:be2+} Linear (top) and non-linear (bottom) spectra of Be$^{2+}$ comparing the exact and the 1D LDA calculation. The inset in the bottom figure shows a zoom into the region from $2.7$ to $3.0$ Ha.}
\end{figure}

\begin{table}
\begin{tabular}{|l|ccccccc||ccc|}\hline
& $\omega_1$ & $\omega_2$ & $\omega_3$ & $\omega_4$ & $\omega_5$ & $\omega_6$ & $\omega_7$ & $\Omega_1$ & $\Omega_2$ & $\Omega_3$\\ \hline
LDA & 1.10 & 1.74 & 1.90 & 1.96 & 2.00 & - & - & 0.22 & 0.40 & -\\
EXX & 1.13 &  1.82 & 2.08 & 2.20 & 2.27 & 2.30 & 2.32 & 0.26 & 0.43 & 0.52\\
exact & 1.12 & 1.81 & 2.08 & 2.19 & 2.26 & 2.29 & 2.32 & 0.28 & 0.42 & 0.54\\
\hline
\end{tabular}
\caption{\label{tab:be2+} Excitation energies from linear and non-linear
response of the 1D Be$^{2+}$ atom corresponding to the spectra in Fig.
\ref{fig:be2+}. Excitations from linear response are denoted as $\omega$ while
those from the non-linear spectrum are denoted with $\Omega$. All numbers are
given in Hartree.}
\end{table}

After having shown that the 1D LDA behaves  very much like its 3D counterpart
for ground-state calculations, we turn our attention to TDDFT where we use it as
an adiabatic approximation to the exact time-dependent exchange-correlation
potential. The propagations were performed in a box ranging from -150 to
150~bohr with absorbing boundary conditions \cite{MCBR2003} and a grid spacing
of 0.2~bohr for a total propagation time of $10^3$~a.u.

In Fig.~\ref{fig:be2+} we compare the spectra calculated in a linear and
non-linear regime from the exact and the LDA calculations for a Be$^{2+}$
system, i.e. a positive charge with $q=4$ and 2 interacting electrons in a
singlet configuration.  In the linear regime a kick of $10^{-4}$~Ha/bohr was
employed at $t=0$ which was then increased to $0.01$~Ha/bohr to obtain the
non-linear response. The values of the excitation energies can be found in
Tab.~\ref{tab:be2+}. In linear response, we see five peaks in the LDA spectrum
which compare well with the first five excitations in the exact case. As
expected, the agreement is better for lower lying excitations and gets worse
the closer we get to the continuum. As a guide for the eye we included the KS
HOMO energy of the LDA calculation and the exact ionization potential. The
onset of the continuum itself appears at too low energies in the LDA
calculation missing two more clearly visible peaks in the exact spectrum. In
other words, the LDA fails to reproduce the proper Rydberg series, a behavior
well known from 3D calculations. For comparison we also included the results
from an EXX calculation which shows a slightly better agreement than LDA for
the first three excitations but, more importantly, reproduces the Rydberg
series due to the correct asymptotic behavior of the corresponding exchange
potential. The quality of the EXX results also implies that correlation is of
secondary importance in the system  for $a=1$. The non-linear spectrum shows
the same excitations as the linear spectrum and three additional peaks for the
exact and the EXX calculation and two additional peaks in the LDA spectrum.
Their energies are also listed in Tab.~\ref{tab:be2+}. Due to the spatial
symmetry of the system all even order responses are zero and the first
non-vanishing higher-order response is of third order. The $\Omega_1=0.28$~Ha
corresponds to an excitation from the second to the third excited state, where
the transition from the ground to the second excited state is dipole forbidden
and, hence, can only be reached in a two-photon process. The other two
frequencies, $\Omega_2=0.42$~Ha and $\Omega_3=0.54$~Ha, correspond to the
transitions from first to second and second to fifth excited state,
respectively. Again, both the EXX and the LDA calculations yield a good
description of the low lying excitations, only the third peak cannot be resolved in the LDA spectrum. 

One feature of the exact spectrum that is missing from both the LDA and the EXX
spectra is the small dip at 2.8~Ha, see inset in Fig.~\ref{fig:be2+}. It results
from a Fano resonance \cite{F1961,FC1968}, i.e. the decay of an excited state
into continuum states. It is missing from both approximate spectra due to the
double-excitation character of the involved excited state. Double excitations
can only be described in TDDFT if a frequency-dependent xc kernel is employed
\cite{MZCB2004}. Any adiabatic approximation, however, leads to a frequency
independent kernel. Hence, double excitations, as well as any resulting
features, are missing from both the ALDA and the AEXX calculations. Apart from
the well-known shortcomings of not including double-excitations and not giving
the correct  Rydberg series, the 1D ALDA reproduces both the linear and the
non-linear exact spectra quite well. 

We have introduced a one-dimensional LDA suitable for the description of
systems interacting via the commonly used soft-Coulomb interaction. We have
shown that the one-dimensional functional is of the same quality as its
three-dimensional counterpart in the calculation of ground-state energies of
atomic systems and the dissociation of small molecules. Also, the linear
spectra show the same quality known from 3D calculations with low energy
excitations being well described while Rydberg and double excitations are
missing. Generally, for the 1D LDA one can expect the same success and failure
in applications that are known from 3D calculations, i.e. the quality of the
LDA results appears to be independent of the dimensionality. We emphasize that
the 1D LDA  yields a good description not only in linear response but also in
the non-linear case. Consequently, one can expect 3D LDA calculations to
perform well for the calculation of non-linear response, where the experimental
data is often difficult to interpret.

The reduced dimensionality of the model systems treated in this work allows for
a direct solution of the interacting problem for small number of particles. The
comparison between the DFT and an exact calculation allows for an assessment of
the quality of approximations beyond what is possible in three-dimensional
systems. One-dimensional model systems can provide useful insight which
hopefully will allow for the construction of new functionals in the future.

We acknowledge funding by the Spanish MEC (FIS2007-65702-C02-01), ACI-promciona
project (ACI2009-1036), ``Grupos Consolidados UPV/EHU del Gobierno Vasco''
(IT-319-07), and the European Community through e-I3 ETSF project (Contract No.
211956).


\begin{thebibliography}{10}

\bibitem{RG1984}
E. Runge and E.~K.~U. Gross, Phys. Rev. Lett. {\bf 52},  997  (1984).

\bibitem{TDDFT2006}
{\em Time-Dependent Density Functional Theory}, Vol.~706 of {\em Lecture Notes
  in Physics}, edited by M.~A.~L. Marques {\it et~al.} (Springer Berlin /
  Heidelberg, 2006).

\bibitem{ORR2002}
G. Onida, L. Reining, and A. Rubio, Rev. Mod. Phys. {\bf 74},  601  (2002).

\bibitem{MBW2002}
N.~T. Maitra, K. Burke, and C. Woodward, Phys. Rev. Lett. {\bf 89},  023002
  (2002).

\bibitem{MZCB2004}
N.~T. Maitra, F. Zhang, R. Cave, and K. Burke, J. Chem. Phys. {\bf 120},  5932
  (2004).

\bibitem{DH2004}
A. Dreuw and M. Head-Gordon, J. Am. Chem. Soc. {\bf 126},  4007  (2004).

\bibitem{M2005}
N. Maitra, J. Chem. Phys. {\bf 122},  234104  (2005).

\bibitem{BWG2005}
K. Burke, J. Werschnik, and E.~K.~U. Gross, J. Chem. Phys. {\bf 123},  062206
  (2005).

\bibitem{VMR2008}
D. Varsano, A. Marini, and A. Rubio, Phys. Rev. Lett. {\bf 101},  133002
  (2008).

\bibitem{CCJ2009}
{\em TDDFT for Molecules and Molecular Solids}, edited by M. Casida, H.
  Chermette, and D. Jacquemin (Special Issue in Journal of Molecular Structure:
  THEOCHEM, vol.{\bf 914}, Issues 1-3 2009).

\bibitem{GMG2009}
L. Goerigk, J. Moellmann, and S. Grimme, Phys. Chem. Chem. Phys. {\bf 11},
  4611  (2009).

\bibitem{GB2009}
O. Gritsenko and E. Baerends, Phys. Chem. Chem. Phys. {\bf 11},  4640  (2009).

\bibitem{SMBWRSD1993}
M. Stahelin {\it et~al.}, J. Chem. Phys. {\bf 98},  5595  (1993).

\bibitem{KDS1998}
P. Kaatz, E. Donley, and D. Shelton, J. Chem. Phys. {\bf 108},  849  (1998).

\bibitem{JES1988}
J. Javanainen, J. Eberly, and Q. Su, Phys. Rev. A {\bf 38},  3430  (1988).

\bibitem{E1990}
J. Eberly, Phys. Rev. A {\bf 42},  5750  (1990).

\bibitem{LGE2000}
M. Lein, E.~K.~U. Gross, and V. Engel, Phys. Rev. Lett. {\bf 85},  4707
  (2000).

\bibitem{L2005}
M. Lein, Phys. Rev. A {\bf 72},  053816  (2005).

\bibitem{TGK2008}
M. Thiele, E.~K.~U. Gross, and S. K\"ummel, Phys. Rev. Lett. {\bf 100},  153004
   (2008).

\bibitem{TMM2009}
D.~G. Tempel, T.~J. Mart\'inez, and N.~T. Maitra, J. Chem. Theory and
  Computation {\bf 5},  770  (2009).

\bibitem{HTR2009}
N. Helbig, I. Tokatly, and A. Rubio, J. Chem. Phys. {\bf 131},  224105  (2009).

\bibitem{LL1977}
{L.D. Landau} and {E.M. Lifschitz}, {\em Quantum Mechanics}
  (Butterworth-Heinemann, Oxford, 1977).

\bibitem{MCBR2003}
M.~A.~L. Marques, A. Castro, G.~F. Bertsch, and A. Rubio, Comp. Phys. Comm.
  {\bf 151},  60  (2003).

\bibitem{CAORALMGR2006}
A. Castro {\it et~al.}, Phys. Stat. Sol. (b) {\bf 243},  2465  (2006).

\bibitem{AMGB2002}
C. Attaccalite, S. Moroni, P. Gori-Giorgi, and G.~B. Bachelet, Phys. Rev. Lett.
  {\bf 88},  256601  (2002).

\bibitem{KS1965}
W. Kohn and L.~J. Sham, Phys. Rev. {\bf 140},  A1133  (1965).

\bibitem{CSS2006}
M. Casula, S. Sorella, and G. Senatore, Phys. Rev. B {\bf 74},  245427  (2006).

\bibitem{SCSM2009}
L. Shulenburger, M. Casula, G. Senatore, and R. Martin, J. Phys. A: Math.
  Theor. {\bf 42},  214021  (2009).

\bibitem{XPTCCR2006}
G. Xianlong {\it et~al.}, Phys. Rev. B {\bf 73},  165120  (2006).

\bibitem{XPAT2006}
G. Xianlong, M. Polini, R. Asgari, and M.~P. Tosi, Phys. Rev. A {\bf 73},
  033609  (2006).

\bibitem{XA2008}
G. Xianlong and R. Asgari, Phys. Rev. A {\bf 77},  033604  (2008).

\bibitem{AB1985b}
C.-O. Almbladh and U. von Barth, Phys. Rev. B {\bf 31},  3231  (1985).

\bibitem{F1961}
U. Fano, Phys. Rev. {\bf 124},  1866  (1961).

\bibitem{FC1968}
U. Fano and J.~W. Cooper, Rev. Mod. Phys. {\bf 40},  441  (1968).

\end{thebibliography}
\end{document}